\documentclass{ws-procs975x65}

\def\lsim{\lower.5ex\hbox{$\; \buildrel < \over \sim \;$}}
\def\gsim{\lower.5ex\hbox{$\; \buildrel > \over \sim \;$}}

\begin{document}

\title{Spectral Signatures of Winds from Accretion Disks Around Black Holes}

\author{S. K. CHAKRABARTI\footnote{\uppercase{A}lso at \uppercase{C}entre for \uppercase{S}pace \uppercase{P}hysics, \uppercase{C}halantika 43, \uppercase{G}aria \uppercase{S}tation \uppercase{R}d.,
\uppercase{K}olkata 700084} ~  and ~ A. NANDI\footnote{\uppercase{W}ork partially
supported by \uppercase{D}epartment of \uppercase{S}cience and \uppercase{T}echnology}}

\address{S.N. Bose National Centre for Basic Sciences,\\
JD-Block, Salt Lake, Kolkata 700098\\
E-mail: chakraba@bose.res.in and anuj@bose.res.in\\ }

\author{A.R. Rao}

\address{Tata Institute of Fundamental Research\\
Homi Bhabha Road, Colaba, Mumbai, 400005\\
E-mail: arrao@tifr.res.in}

\maketitle

\abstracts{We show that with the wind/jet activity,  
the spectral index of hard X-ray is changed in galactic microquasars.
When mass loss takes place, the spectrum becomes softer and when 
mass gain takes place, the spectrum becomes harder. We present 
examples of such changes in GRS1915+105.}

\noindent To be Published in the proceedings of the 10th Marcel Grossman Meeting (World Scientific Co., Singapore),
Ed. R. Ruffini et al.

\section{Introduction}

Accretion and outflows are common in systems which
include compact objects such as black holes and neutron stars. 
In the last three decades, numerous models have been
built to produce jets and outflows. In the case of black holes, since
no matter can escape out of the horizon, some suggested that the jets 
are formed in the strong field region around black holes in the form of pair plasma (Type A) while others suggested that
the jets come from the accretion disks (Type B). Models of Type A are still incomplete and
most likely to be incorrect since jets are very generic, and equally energetic 
jets are also possible for neutron stars as well. Until 1998, all the 
Type B models envisaged that the {\it whole} accretion disk participates in the jet formation.
In 1998, Chakrabarti suggested that because the centrifugal barrier dominated boundary layer
(CENBOL) in an accretion flow is hot (being the post-shock region), it is more
likely that jets would be produced from this region (Chakrabarti, 1998; 1999a). 
Outflow rate in units of inflow rate was also computed successfully. 
There are strong observational supports of the original assertion that the winds do indeed  
originate from the CENBOL area of an accretion flow. Subsequently, Chakrabarti (1999b),
showed that when matter leaves the CENBOL in the form of an outflow, it reduces the electron number
density in the CENBOL and clearly softens the emitted spectrum. 
Chakrabarti et al. (2002) showed that not only does the hard spectrum is softened
when the outflow is ON, a soft spectrum is hardened when the outflow is cooled down by external
photons and `fall back' onto the CENBOL area. Thus quasi-periodic burst-on and burst-off states
observed in the black hole GRS 1915+105 cause quasi-periodic softening and hardening of spectrum.
In the following, we give some of the examples of such results.  

It is to be noted that the radiation from the wind itself is immaterial, since its
emitting volume is small. The only significant effect that a wind can have is the 
modification of electron number density in the CENBOL. This is what is highlighted 
in the present study.  Alternatively, one could imagine that the CENBOL itself is the
base of the jet though results remain the same as ours.

\section{Theoretical Background}

What could be the spectral signature of a CENBOL which is losing matter
to hot winds or gaining matter from the cooler winds? The two component advective flow
(TCAF) model of Chakrabarti \& Titarchuk (1995) and Chakrabarti (1997) give some insight into these problems.
In Fig. 1, we present two sets of calculated spectra using the following parameters:
disk accretion rates for the soft and the hard states are $0.1 {\dot M}_{Edd}$
and $0.02 {\dot M}_{Edd}$ respectively and the halo accretion rate is kept fixed at
$0.2 {\dot M}_{Edd}$. The shock is located at $X_{s}=6$ and $X_s=40$
respectively in high/soft and low/hard states (dotted curves) respectively. 
This is in line with the general conclusions that the inner edge moves in during 
soft states. The solid curves are computed results which mimic the burst-off (mass-loss) and burst-on

\begin{figure}[hp] 
\vskip -4.0cm
\centerline{\epsfxsize=4.1in\epsfbox{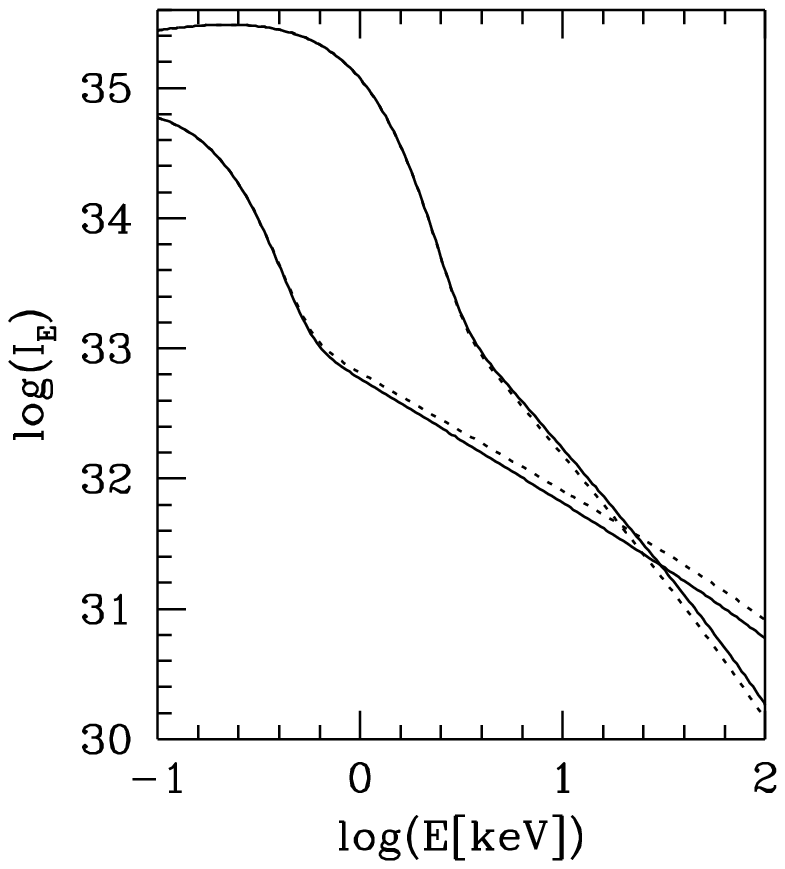}}   
\vskip -1.0cm
{\small{\bf Fig. 1:} Calculated spectra (uncorrected for absorption)
for burst-off and burst-on states (solid) and low and high states (dotted).
In burst-off state, spectrum appears to be softened with respect to low state, and in burst-on state,
spectrum appear to be hardened with respect to the high state. As a result the pivoting point shifts towards higher energy. }
\end{figure}

\noindent (mass-gain) state spectra. In these
cases, the disk accretion rates are kept as before, but twenty
percent of CENBOL matter is assumed to be lost in wind in the burst-off
state (Molteni, Lanzafame \& Chakrabarti, 1994)
and ten percent (for example) of matter in outflow is assumed to be
falling back on the halo in the burst-on state.
Because of the selective softening and hardening, the intersection (pivot) point
of the resulting spectra is located at a higher energy.

\section{Observational Results}

The black hole candidate GRS~1915+105 was in a low state from December, 1996 up to April, 1997.
The spectral and temporal behavior during the low state was stable
characterized by a hard spectrum (with the power-law photon index
of $\sim$2, and the total flux in the power-law component being
$\sim$80\%) and $1-10$ Hz QPOs (Trudolyubov et al.\ 1999; Muno et al. \ 1999).
Since the source was in a similar state on several
occasions (1996 July-August; 1997 October; 1998 September-October),
we treat this state as the low state of GRS~1915+105. The source reached
a high state in 1997 August. We have selected one observation
each from the low state (1997 March 26 with PID number of 20402-01-21-00;
Muno et al. 1999 and Trudolyubov et al.\ 1999)
and high state (1997 August 19 with a PID number of 20402-01-41-00; Muno et al.\  1999) to quantify
the spectral parameters. We present the spectra of the source during the irregular
bursts (Yadav et al. 1999) observed on 1997 June 18 (PID 20402-01-33-00) in Fig. 2a,
on 1997 July 10 (PID 20402-01-36-00) in Fig. 2b and on 1997 July 12
(PID 20402-01-37-01) in Fig. 2c.  

\begin{figure}[hp] 
\vskip -0.8cm
\centerline{\epsfxsize=4.7in\epsfbox{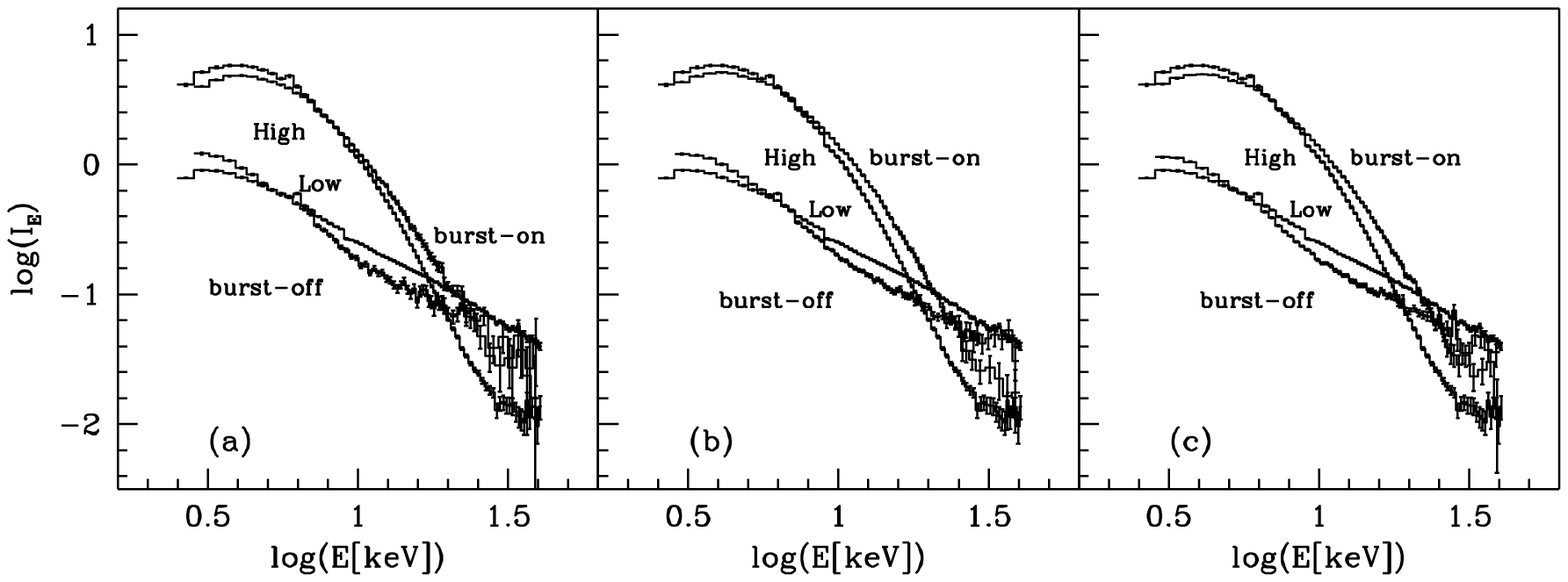}}   
\vskip -6.5cm
{\small{\bf Fig. 2:} 
The unfolded RXTE-PCA spectra of GRS~1915+105 obtained during the low and high states
are compared with burst-on and burst-off spectra during the irregular bursts observed on (a) 1997 June 18,
(b) 1997 July 10, and (c) 1997 July 12. The histogram shows
the fitted model. During the burst-off/burst-on transition, the
pivoting occurs at a higher energy.}
\end{figure}

It is seen that in the low state, the disk blackbody component has
lower temperature (kT$_{in}$ = 0.60$\pm$0.05 keV) and a larger inner disk
radius (R$_{in}$ = 115$\pm$2 km) compared to the high state, which has
the inner disk temperature of 1.95$\pm$0.01 keV and the inner disk radius
of 26$\pm$1 km. The inner disk parameters are calculated using a distance
to the source of $10$ kpc and inclination to the disk of 70$^\circ$.
The power-law index in the low state (2.40$\pm$0.01) is much flatter
than that seen in the high state (3.61$\pm$0.02). 
The disk blackbody component has a 3 $-$ 26 keV
flux of $<$10\% of the total flux in the low state, which increase to $>$65\%.
The two spectra intersect at around $17$ keV.

The corresponding unfolded spectra during the irregular burst of June 18, 1997
presented in Figure 2a shows that the burst-off state has spectral parameters
similar to that of the low  state (kT$_{in}$ = 0.76 keV, index = 2.76)
and the burst-on state spectrum resembles that of the high state
(kT$_{in}$ = 2.2  keV, index = 3.1). It is clear
that the spectrum of the burst-off state is softer than that
of the low state and the spectrum of the burst-on state is harder
than that of the high/soft state. As a result, the energy at which
the two spectra pivot is much higher ($\sim$ 26 keV). The same behaviour is
seen also in Fig. 2b and 2c where the IDs used were 20402-01-36-00 and 20402-01-37-01.
The low and high states have been kept as above. The intersection in all these cases
during the burst-on/burst-off transition is around 25 keV ---
far above the low/high intersection.

\section{Discussion and conclusions}

A comparison of the two spectral states in Fig. 2(a-c) shows that
in GRS1915+105 during the  bursting phase, the low energy component ($\lsim$ 5 keV) remains essentially
the same as found during the extended low and high states of the
source. The major change is above $\sim$ 10 keV where the hard component
of the low state softens to form a burst-off state and that of the high state hardens to form
a burst-on state. These we interpret to be the signatures of the wind activities
in an accretion flow.

\section*{Acknowledgments} 
AN acknowledges DST project grant No.  SP/S2/K-15/2001 for funding his research.

\end{document}